\documentclass[12pt]{article}

\usepackage{geometry}
\usepackage{enumitem}
\setlist{nolistsep}
\usepackage[T1]{fontenc}
\usepackage{lmodern}
\usepackage{mathtools}

\usepackage[backend=bibtex,style=numeric-comp]{biblatex}
\bibliography{bibtex}

\usepackage[hidelinks]{hyperref}
\usepackage[frozencache=true,cachedir=minted-cache]{minted}
\usepackage{pgfplots}
\pgfplotsset{compat = newest}
\usepackage{graphicx}
\usepackage{array}
\usepackage{caption}
\usepackage{subcaption}

\newcommand{\im}[1]{\includegraphics[scale=0.28]{img/#1.png}}

\begin{document}
	
\begin{center}
\LARGE \textbf{Generalized Projection Matrices}
\end{center}

\centerline{S.J.D. MacIntosh}

\vspace{2em}

\centerline{\begin{minipage}{0.9\textwidth}
Projection matrices are necessary for a large portion of rendering computer graphics. There are primarily two different types of projection matrices -- perspective and orthographic -- which are used frequently, and are traditionally treated as mutually incompatible with each other in how they are defined. Here, we bridge the gap between the two different forms of projection matrices to present a single generalized projection matrix that can represent both.

\vspace{0.5em}

\end{minipage}}
	

\section{Introduction}

In graphics programming, projection matrices are very common. They serve as a method to convert eye coordinates into clip coordinates, which allows for easy conversion into normalized device coordinates and ultimately rendering to a screen \autocite{cgpp}.

The two primary forms of projection matrices are \textit{perspective} or \textit{orthographic}. A simple scene rendering with perspective projection is pictured left, while the same scene rendered with orthographic projection is pictured right. Note that the checkerboard plane is not visible in the orthographic projection because it is perfectly perpendicular to the direction of the projection.

\vspace{1em}
\begin{figure}[h]
\centering
\begin{subfigure}[b]{0.49\textwidth}
	\centering
	\includegraphics[scale=0.49]{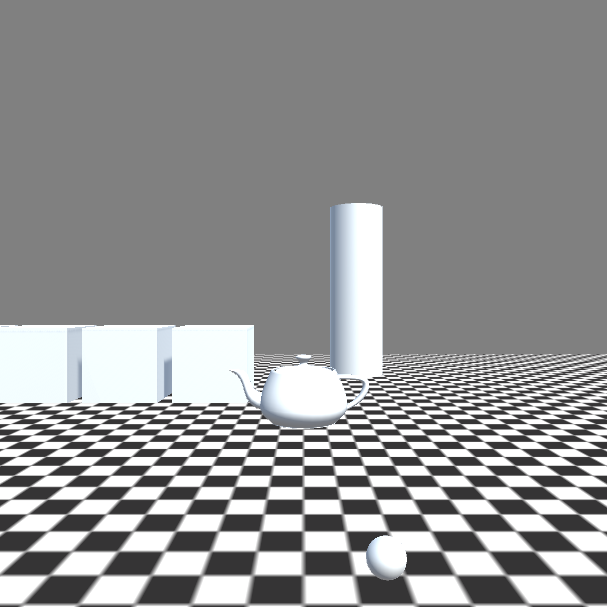}
	\caption{Perspective projection}
\end{subfigure}
\hfill
\begin{subfigure}[b]{0.49\textwidth}
	\centering
	\includegraphics[scale=0.49]{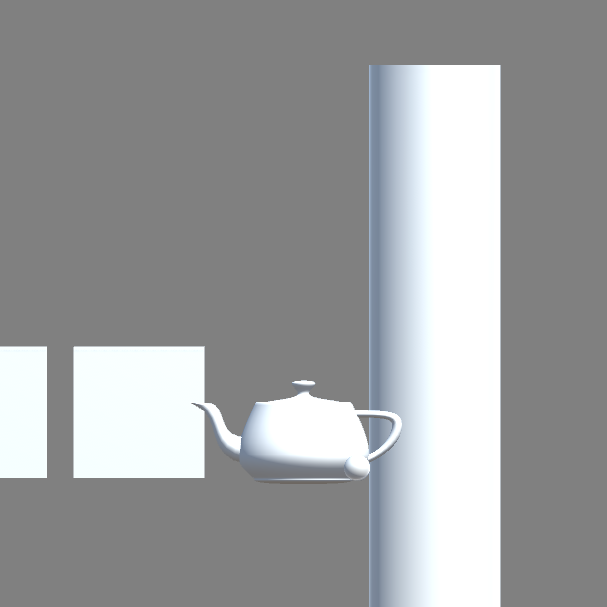}
	\caption{Orthographic projection}
\end{subfigure}
\caption{A comparison of a scene rendered under different projections.}
\label{fig:three}
\end{figure}

The most visible difference is that perspective projections factor in distance (as in, objects that are further away from the camera will appear smaller), while orthographic projections have things appear the same size regardless of distance.

Both see frequent use in computer rendering. Perspective projection is used primarily for 3D scenes because it transforms coordinates in a manner more visually familiar to us. Orthographic projection, in comparison, is more frequently used for 2D scenes (or 2D elements within an otherwise 3D scene, such as rendering the heads-up display in a videogame) as well as isometric visuals (and related concepts like dimetric or trimetric projection). Beyond this, orthographic projections are used in situations such as engineering \autocite{tpcg}.

Projection matrices can be represented by a frustum. Drawing the frustum allows us to see what is within the boundaries of the projection and will become visible. Similar to the previous example, on the left is perspective projection and on the right is orthographic projection:

\vspace{1em}
\begin{figure}[h]
	\centering
	\begin{subfigure}[b]{0.49\textwidth}
		\centering
		\includegraphics[scale=0.49]{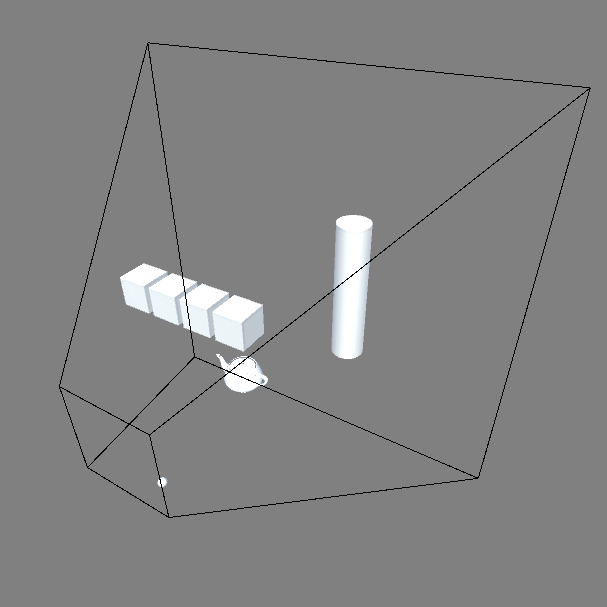}
		\caption{Perspective projection}
	\end{subfigure}
	\hfill
	\begin{subfigure}[b]{0.49\textwidth}
		\centering
		\includegraphics[scale=0.49]{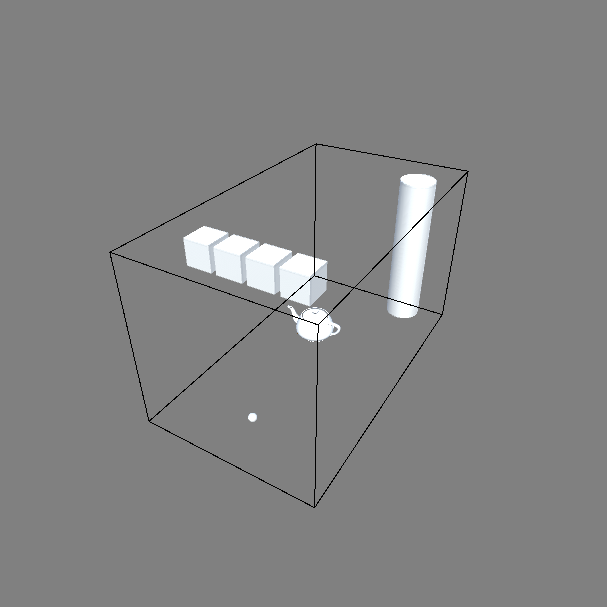}
		\caption{Orthographic projection}
	\end{subfigure}
	\caption{A comparison of frustums associated with different projections.}
	\label{fig:three}
\end{figure}

These two types of projection are generally treated as distinct, meaning that a projection matrix can either be perspective or it can be orthographic, and the mathematics for each are determined independently. When a programmer uses a graphics API they can expect to see one function to generate a perspective projection matrix, and then a different function to generate an orthographic projection matrix.

This is sufficient for many uses, but it can also be limiting. Keeping them distinct means one cannot smoothly interpolate between a perspective and an orthographic projection. Neither could one apply some subtle orthographic projection to an otherwise perspective projection (or vice versa) as a stylistic choice. This will be the focus of this paper.

This idea is not strictly new -- some software such as the Unity Engine will do something similar in specific circumstances \autocite{unity2d3dToggle}, and we have previously published code that follows a similar technique as we will describe \autocite{beowolf}. However, we are not aware of any descriptions for its design, and we offer additional features on top of this to make it sufficiently novel. Although it is niche, we have described use cases above for a combination of orthographic and perspective projection together into a single matrix or programming function.

\section{Partially-Orthographic Projection}

Traditionally, using OpenGL as an example, a perspective projection matrix is defined in a manner such as \autocite{gluPerspective, gluPerspectiveCode}:

\[
P_p =
\left[ {\begin{array}{cccc}
		\frac{\cot(\theta / 2)}{\alpha} & 0 & 0 & 0\\
		0 & \cot(\theta / 2) & 0 & 0\\
		0 & 0 & -\frac{f + n}{f - n} & \frac{-2fn}{n-f}\\
		0 & 0 & -1 & 0\\
\end{array} } \right]
\]

Where $\theta$ is the vertical angular field of view (FOV), $\alpha$ is the aspect ratio, $f$ is the far plane distance, and $n$ is the near plane distance. Other graphics APIs may differ from OpenGL, but the concept is shared between them.

Similarly, in OpenGL an orthographic projection can be defined as \autocite{glOrtho}:

\[
P_o =
\left[ {\begin{array}{cccc}
		\frac{1}{r} & 0 & 0 & 0\\
		0 & \frac{1}{t} & 0 & 0\\
		0 & 0 & \frac{-2}{f - n} & -\frac{f+n}{f-n}\\
		0 & 0 & 0 & 1\\
\end{array} } \right]
\]

Where $r$ is the distance from the center to the right edge of the frustum, and $t$ is the distance from the center to the top edge of the frustum.

A partially-orthographic projection, which is somewhere between perspective and orthographic projections will now be outlined. Consider a variable $p$ in which a value of $0$ means that we use perspective projection, a value of $1$ means we use orthographic projection, a value of $0.5$ means we use a 50\%-50\% mix of both, and so on.

This effect can be achieved through component-wise interpolation between the perspective projection matrix and the orthographic projection matrix. This results in the following matrix:

\[
P_{c1} =
\left[ {\begin{array}{cccc}
		l(\frac{\cot(\theta / 2)}{\alpha}, \frac{1}{r}, p) & 0 & 0 & 0\\
		0 & l(\cot(\theta / 2),\frac{1}{t},p) & 0 & 0\\
		0 & 0 & l(-\frac{f + n}{f - n},\frac{-2}{f - n},p) & l(\frac{-2fn}{n-f},-\frac{f+n}{f-n},p)\\
		0 & 0 & p-1 & p\\
\end{array} } \right]
\]

Where $l$ is the linear interpolation function:

\[l(x,y,z) = x(1-z) + yz\]

In terms of code, assuming that we have the necessary perspective and orthographic functions to obtain their respective projection matrices, it can be written as follows:

\begin{minted}[fontsize=\footnotesize,obeytabs=true,tabsize=2,breaklines]{csharp}
Matrix4x4 GeneralProjectionMatrix(float theta, float alpha, float r, float t, float n, float f, float p)
{
	Matrix4x4 perspective = Matrix4x4.Perspective(theta, alpha, n, f);
	Matrix4x4 orthographic = Matrix4x4.Ortho(-r, r, -t, t, n, f);
	
	// combine matrices
	Matrix4x4 combined;
	for (int i = 0; i < 4; i++)
		for (int j = 0; j < 4; j++)
			combined[i,j] = (1-p) * perspective[i,j] + p * orthographic[i,j];
		
	return combined;
}
\end{minted}

This is not the only way to write this function, it is presented this way for readability purposes. Another way would be to set each component of the matrix directly, according to the projection matrix derived above. 

\section{FOV-Distance Equivalence}

At this point, the orthographic projection in the model is missing the ability to define itself fully in terms of the perspective projection. The relation between $r$ and $t$ used by the orthographic projection and $\theta$ and $\alpha$ of the perspective projection is not clear. Both deal with similar concepts -- the horizontal and vertical sizes of the projection matrix -- but it is not easy to see what values they should be.

Here we add a value $d$ that represents the distance for which the two forms of projection have the same FOV, which will allow us to derive $r$ and $t$ from $\theta$ and $\alpha$.

The equation to find the height of the projection frustum at a particular distance is given by $h = 2\mathrm{tan}(\theta / 2)d$ where $d$ is the distance we are considering.

$r$ and $t$ can now be defined as:

\[t = \tan(\theta / 2)d\]
\[r = \alpha t\]

We can then apply this directly into our projection matrix:

\[
P_{c2} =
\left[ {\begin{array}{cccc}
		l(\frac{\cot(\theta / 2)}{\alpha}, \frac{1}{\alpha\tan(\theta / 2)d}, p) & 0 & 0 & 0\\
		0 & l(\cot(\theta / 2),\frac{1}{\tan(\theta / 2)d},p) & 0 & 0\\
		0 & 0 & l(-\frac{f + n}{f - n},\frac{-2}{f - n},p) & l(\frac{-2fn}{n-f},-\frac{f+n}{f-n},p)\\
		0 & 0 & p-1 & p\\
\end{array} } \right]
\]

And add to our code:

\begin{minted}[fontsize=\footnotesize,obeytabs=true,tabsize=2,breaklines]{csharp}
Matrix4x4 GeneralProjectionMatrix(float theta, float alpha, float n, float f, float p, float d)
{
	Matrix4x4 perspective = Matrix4x4.Perspective(theta, alpha, n, f);
	
	float t = Mathf.Tan(theta / 2.0f) * d;
	float r = t * alpha;
	Matrix4x4 orthographic = Matrix4x4.Ortho(-r, r, -t, t, n, f);
	
	// combine matrices
	Matrix4x4 combined;
		for (int i = 0; i < 4; i++)
			for (int j = 0; j < 4; j++)
				combined[i,j] = (1-p) * perspective[i,j] + p * orthographic[i,j];
	
	return combined;
}
\end{minted}

\section{Shear}

In projection matrices, shear refers to the amount that the far plane shifts up, down, left, or right relative to the forward direction of the camera. This can produce a skewed projection matrix, with different effects for the type of projection matrix being used. For perspective matrices, this can be used for purposes such as stereoscopic images \autocite{Kooima2009GeneralizedPP}. With orthographic matrices, this results in oblique projections \autocite{tpcg}.

With a perspective matrix, vertical shear can be achieved by changing the $P_{1,2}$ component in the matrix. Similarly, horizontal shear is dependent on the $P_{0,2}$ component.

A value of $1$ for vertical shear on a projection matrix will shift the far face of the frustum upwards by half its height. $-1$ does the same in the opposite direction. For horizontal shear, $1$ skews to the right while $-1$ skews towards the left.

Orthographic matrices work similarly, except with a conversion factor of $1/d$ for the center of the far face to end up in the same spot as with perspective projection. In other words orthographic projection matrices need to have their shear factors divided by the distance for which our FOV is equivalent between the two forms of projection.

We can augment our generalized projection matrix with this addition:

\[
P_{c3} =
\left[ {\begin{array}{cccc}
		l(\frac{\cot(\theta / 2)}{\alpha}, \frac{1}{\alpha\tan(\theta / 2)d}, p) & 0 & l(s_h, s_h/d, p) & 0\\
		0 & l(\cot(\theta / 2),\frac{1}{\tan(\theta / 2)d},p) & l(s_v, s_v/d, p) & 0\\
		0 & 0 & l(-\frac{f + n}{f - n},\frac{-2}{f - n},p) & l(\frac{-2fn}{n-f},-\frac{f+n}{f-n},p)\\
		0 & 0 & p-1 & p\\
\end{array} } \right]
\]

Where $s_h$ is the horizontal shear factor and $s_v$ is the vertical equivalent.

We can then add this capability to our code by setting these components after generating the rest of the matrix:

\begin{minted}[fontsize=\footnotesize,obeytabs=true,tabsize=2,breaklines]{csharp}
Matrix4x4 GeneralProjectionMatrix(float theta, float alpha, float n, float f, float p, float d, float sh, float sv)
{
	Matrix4x4 perspective = Matrix4x4.Perspective(theta, alpha, n, f);
	
	float t = Mathf.Tan(theta / 2.0f) * d;
	float r = t * alpha;
	Matrix4x4 orthographic = Matrix4x4.Ortho(-r, r, -t, t, n, f);
	
	// combine matrices
	Matrix4x4 combined;
	for (int i = 0; i < 4; i++)
		for (int j = 0; j < 4; j++)
			combined[i,j] = (1-p) * perspective[i,j] + p * orthographic[i,j];
			
	// apply shearing
	combined[0,2] = (1-p) * sh + p * (sh / d);
	combined[1,2] = (1-p) * sv + p * (sv / d);
	
	return combined;
}
\end{minted}

\section{Variations}

The projection matrix and code we have described so far is now usable. However, one might prefer a different way to define an equivalent projection matrix, or supplement what we have with additional features. Several alternatives or optional suggestions are outlined below.

\subsection{Vertical FOV + Horizontal FOV}

Currently the horizontal FOV is implicit, equal to the vertical FOV multiplied $\theta$ by the aspect ratio $\alpha$.

The horizontal FOV  can be made explicit and the aspect ratio implicit, by calculating $\alpha = \theta' / \theta$ where $\theta'$ is the horizontal FOV. We can then continue as before.

Provided is an example code wrapper for this function:

\begin{minted}[fontsize=\footnotesize,obeytabs=true,tabsize=2,breaklines]{csharp}
Matrix4x4 GeneralProjectionMatrix(float theta, float thetaTick, float n, float f, float p, float d, float sh, float sv)
{
	float alpha = (thetaTick / theta);
	return GeneralProjectionMatrix(theta, alpha, n, f, p, d, sh, sv);
}
\end{minted}

\subsection{Horizontal FOV + Aspect Ratio}

An alternative to defining the vertical FOV $\theta$ and using the aspect ratio $\alpha$ would be to instead use the horizontal FOV $\theta'$ alongside $\alpha$. This way, we derive $\theta$ from $\theta'$ and $\alpha$ by the equation $\theta = \theta' / \alpha$.

Like above, we provide an example wrapper that performs this conversion:

\begin{minted}[fontsize=\footnotesize,obeytabs=true,tabsize=2,breaklines]{csharp}
Matrix4x4 GeneralProjectionMatrix(float thetaTick, float alpha, float n, float f, float p, float d, float sh, float sv)
{
	float theta = (thetaTick / alpha);
	return GeneralProjectionMatrix(theta, alpha, n, f, p, d, sh, sv);
}
\end{minted}

\subsection{Percentage Tweaking}

The partially-orthographic percentage may be unintuitive; while it makes sense that $p = 0$ is fully perspective and $p = 1$ is fully orthographic, a value of $p = 0.5$ may not visually appear to be an equal mix of both.

Let us denote a mapping function $m(x)$. There are constraints on what our mapping function can be, namely that $m(0) = 0$ and $m(1) = 1$. These constraints are easy to satisfy and there are many candidate functions that meet our needs.

One example mapping function would be $m(x) = x^{1/c}$ where $c$ is a positive constant. Here are some examples with various values for $c$:

\centerline{
\begin{tikzpicture}
\begin{axis}[xmin = 0, xmax = 1,ymin = 0, ymax = 1, legend pos=south east]
\addplot[domain = 0:1,samples = 200,smooth,thick,black] {x};
\addplot[domain = 0:1,samples = 200,smooth,thick,blue] {x^(1/3)};
\addplot[domain = 0:1,samples = 200,smooth,thick,green] {x^(1/5)};
\addplot[domain = 0:1,samples = 200,smooth,thick,yellow] {x^(1/7)};
\addplot[domain = 0:1,samples = 200,smooth,thick,red] {x^(1/9)};

\addlegendentry{$x$}
\addlegendentry{$x^{1/3}$}
\addlegendentry{$x^{1/5}$}
\addlegendentry{$x^{1/7}$}
\addlegendentry{$x^{1/9}$}
\end{axis}
\end{tikzpicture}
}

\begin{figure}[h]
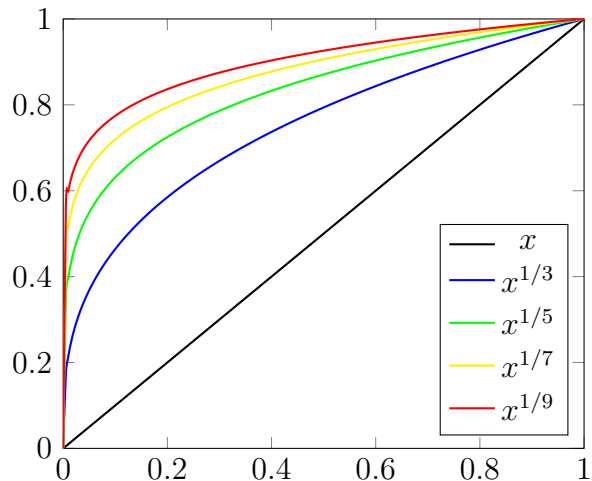

\caption{Different mapping functions compared over the range $[0,1]$}
\end{figure}

Appendix A offers a demonstration and comparison of these mapping functions with various values of $p$.

\subsection{Infinite Projection Matrix}

A technique that sees some use is known as the \textit{infinite projection matrix}, where the frustum extends infinitely far. This allows for rendering at any distance without having to worry about the far plane.

We can generate an infinite projection matrix by taking the limit of our projection matrix as $f$ goes to infinity \autocite{InfProjMatrix}:

\[
P_{i1} = \left[ {\begin{array}{cccc}
		l(\frac{\cot(\theta / 2)}{\alpha}, \frac{1}{\alpha\tan(\theta / 2)d}, p) & 0 & l(s_h, s_h/d, p) & 0\\
		0 & l(\cot(\theta / 2),\frac{1}{\tan(\theta / 2)d},p) & l(s_v, s_v/d, p) & 0\\
		0 & 0 & l(-1,0,p) & l(-2n,-1,p)\\
		0 & 0 & p-1 & p\\
\end{array} } \right]
\]

If there are precision issues as a result of this, we can tweak these values with a small $\epsilon$ \autocite{InfProjMatrix}:

\[
P_{i2} = \left[ {\begin{array}{cccc}
		l(\frac{\cot(\theta / 2)}{\alpha}, \frac{1}{\alpha\tan(\theta / 2)d}, p) & 0 & l(s_h, s_h/d, p) & 0\\
		0 & l(\cot(\theta / 2),\frac{1}{\tan(\theta / 2)d},p) & l(s_v, s_v/d, p) & 0\\
		0 & 0 & l(\epsilon-1,0,p) & l((\epsilon-2)n,\epsilon-1,p)\\
		0 & 0 & p-1 & p\\
\end{array} } \right]
\]

Similar to our approach for shearing, we can easily add this to our existing code by setting these components manually:

\begin{minted}[fontsize=\footnotesize,obeytabs=true,tabsize=2,breaklines]{csharp}
// if `f == -1`, use an infinite projection matrix
Matrix4x4 GeneralProjectionMatrix(float theta, float alpha, float n, float f, float p, float d, float sh, float sv, float epsilon)
{
	Matrix4x4 perspective = Matrix4x4.Perspective(theta, alpha, n, f);
	
	float t = Mathf.Tan(theta / 2.0f) * d;
	float r = t * alpha;
	Matrix4x4 orthographic = Matrix4x4.Ortho(-r, r, -t, t, n, f);
	
	// combine matrices
	Matrix4x4 combined;
	for (int i = 0; i < 4; i++)
		for (int j = 0; j < 4; j++)
			combined[i,j] = (1-p) * perspective[i,j] + p * orthographic[i,j];
	
	// apply shearing
	combined[0,2] = (1-p) * sh + p * (sh / d);
	combined[1,2] = (1-p) * sv + p * (sv / d);
	
	// apply infinite projection
	if (f == -1)
	{
		combined[2,2] = (1-p) * (epsilon - 1);
		combined[2,3] = (1-p) * ()(epsilon - 2) * n) + p * (epsilon - 1);
	}
	
	return combined;
}
\end{minted}

\subsection{Value Enforcement}

Not every value for every variable is supported in this projection matrix -- for example, $\alpha = 0$ results in a division by 0, among other division by 0 issues. With others, such as $d < 0$, it may not be as apparent that it will cause issues, however this can be quickly verified as an issue by testing a projection matrix with these values.

We present some sensible restrictions for the values listed:

\begin{itemize}
	\item $0 < \theta < \pi$ using radians, $0 < \theta < 180$ using degrees
	\item $0 < \alpha$
	\item $0 \le p \le 1$
	\item $0 < d$. A user may find $n \le d \le f$ to be more intuitive, but there's no reason that $d$ has to exist in the range $[n,f]$.
	\item $0 < f$, except for the case of $f = -1$ for infinite projection matrices.
	\item $0 < n < f$, when $f \ne -1$. If $f = -1$, $0 < n$.
	\item $0 < \epsilon$, where $\epsilon$ should be very small. Lengyel \autocite{InfProjMatrix} recommends $|\epsilon| \ge 2^{-21} \approx 4.8 \times 10^{-7}$.
\end{itemize}

\section{Conclusion}

We set out to create a unified projection matrix that would allow us to use both perspective and orthographic projection by changing a single value. We have also discussed multiple variants and additional features for this generalized projection matrix. Both the generalized projection matrix, as well as code to generate it, have been offered. The end result is the ability to blend between perspective and orthographic projection, with more advanced features such as shearing, via an intuitive API.

\subsection{Acknowledgements}

Many thanks to my partner, Dr. Carly Lilley, DVM, for her extensive feedback and support.

\newpage

\printbibliography

\newpage

\section*{Appendix A}

\begin{center}
	\begin{tabular}{ >{\centering\arraybackslash}m{.08\linewidth} >{\centering\arraybackslash}m{\widthof{\im{scene_0}}} >{\centering\arraybackslash}m{\widthof{\im{scene_0}}}>{\centering\arraybackslash}m{\widthof{\im{scene_0}}} }
		& $p = 0.25$ & $p = 0.5$ & $p = 0.75$ \\ \\
		$x$ & \im{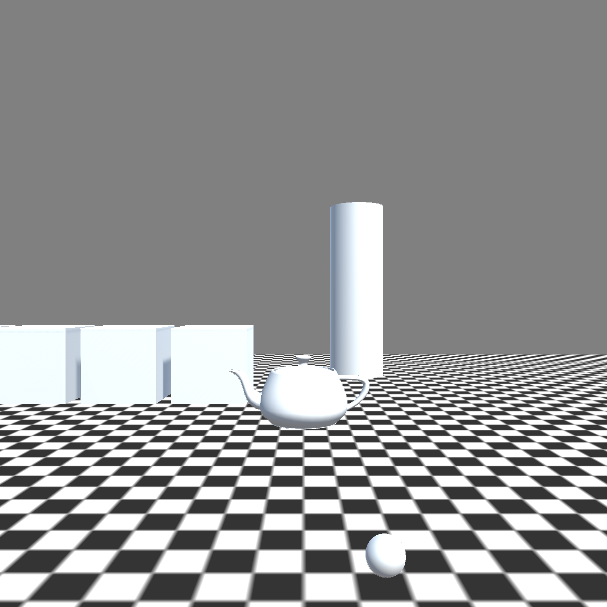} & \im{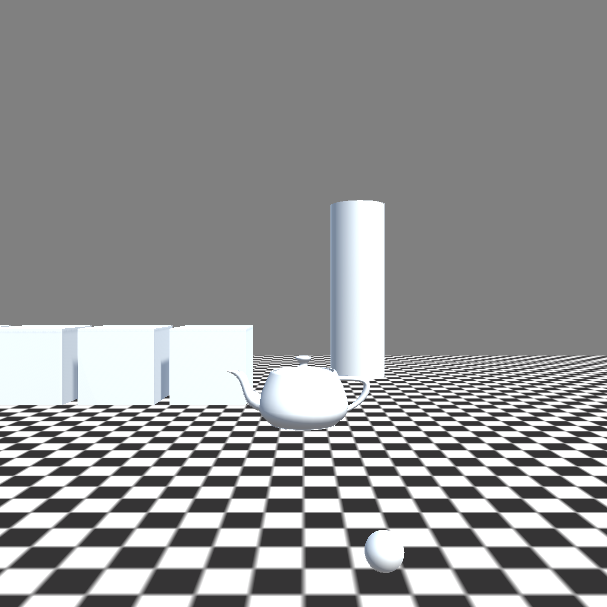} & \im{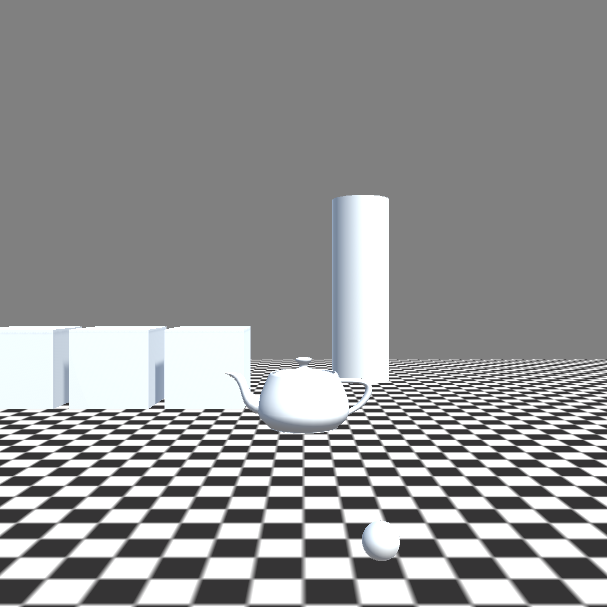} \\ 
		$x^{(1/3)}$ & \im{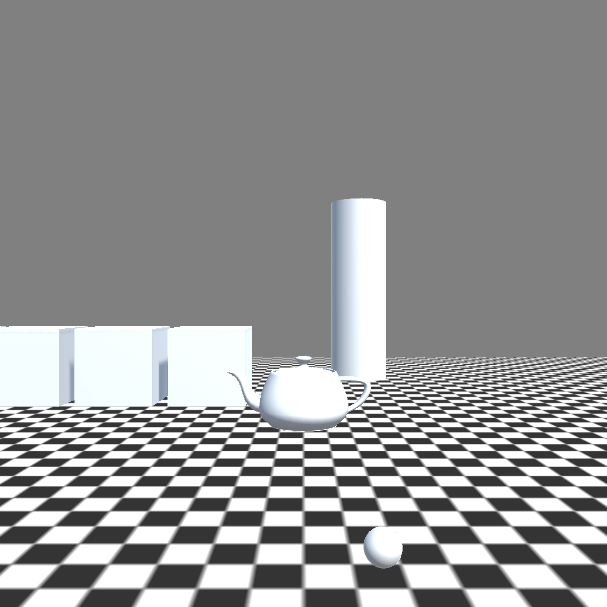} & \im{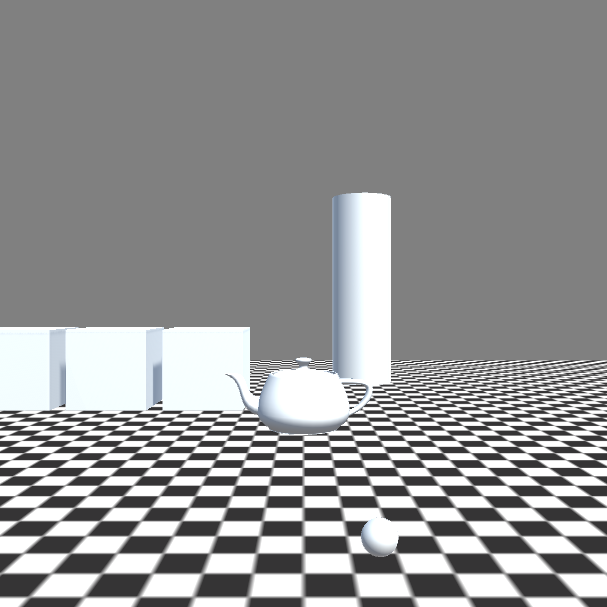} & \im{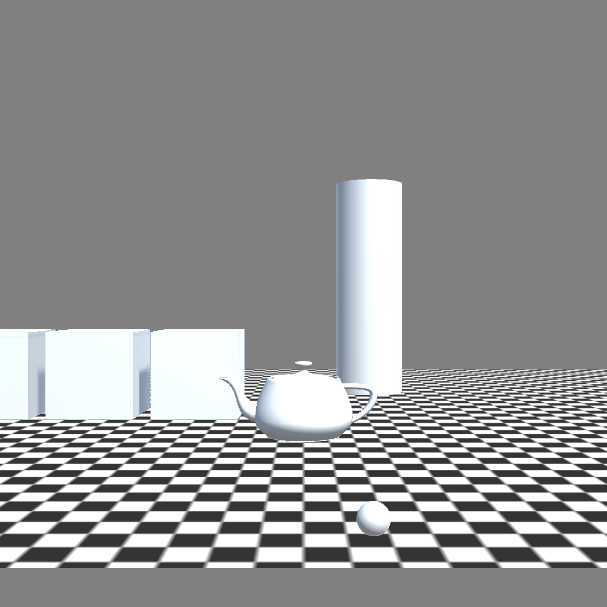} \\ 
		$x^{(1/5)}$ & \im{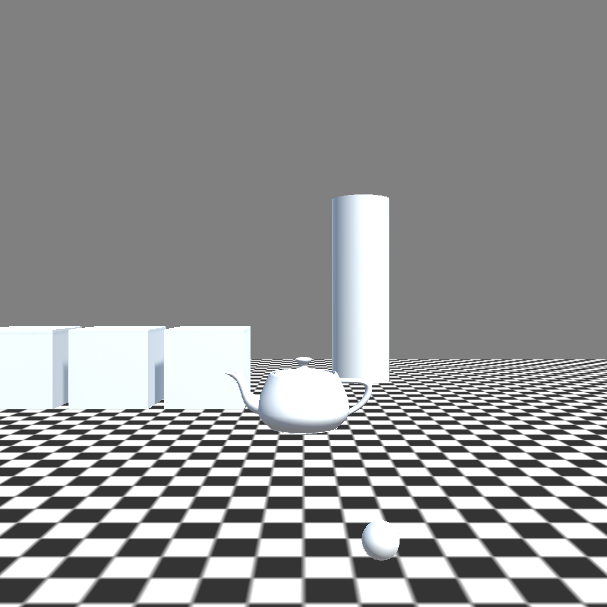} & \im{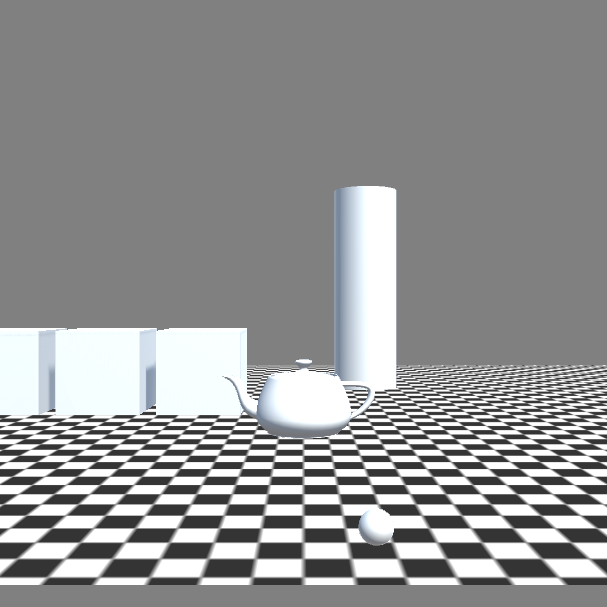} & \im{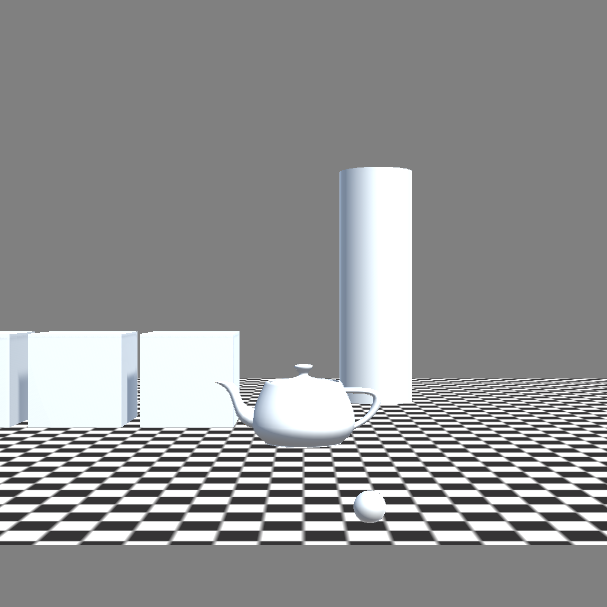} \\ 
		$x^{(1/7)}$ & \im{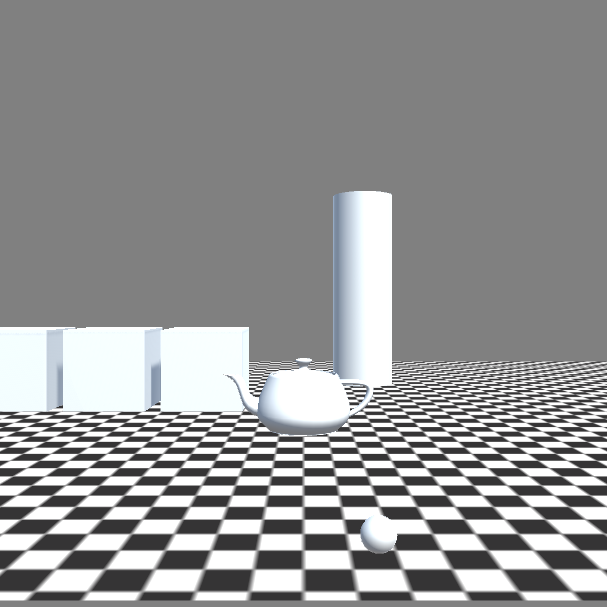} & \im{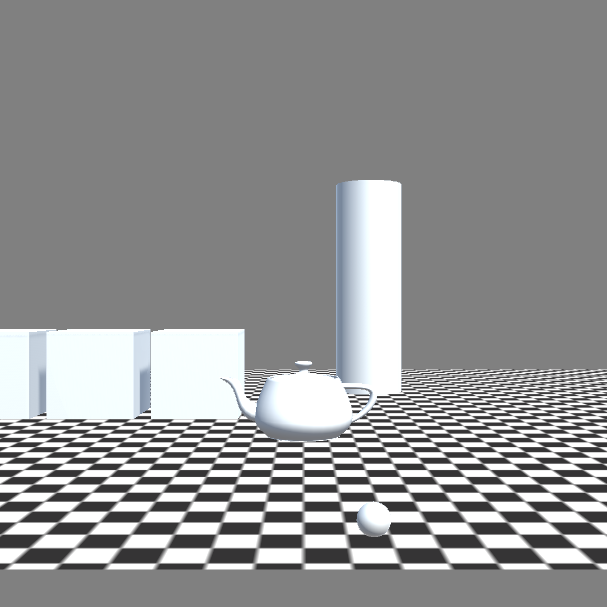} & \im{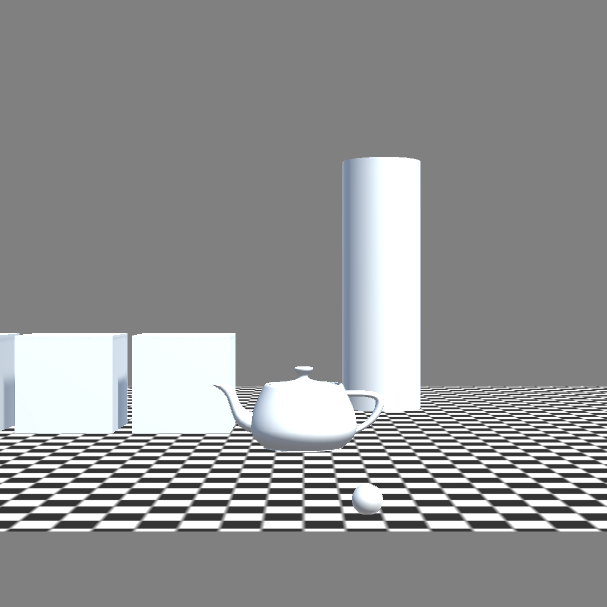} \\ 
		$x^{(1/9)}$ & \im{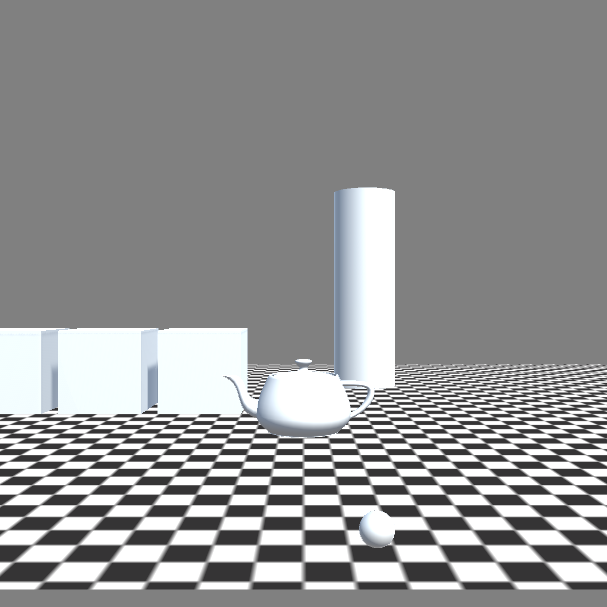} & \im{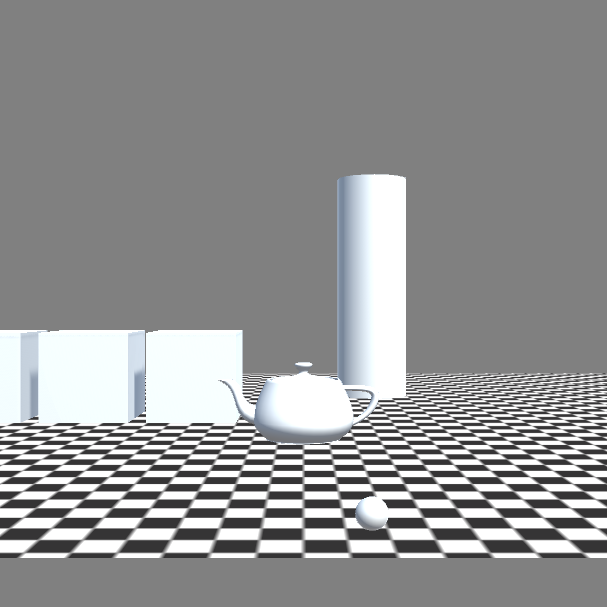} & \im{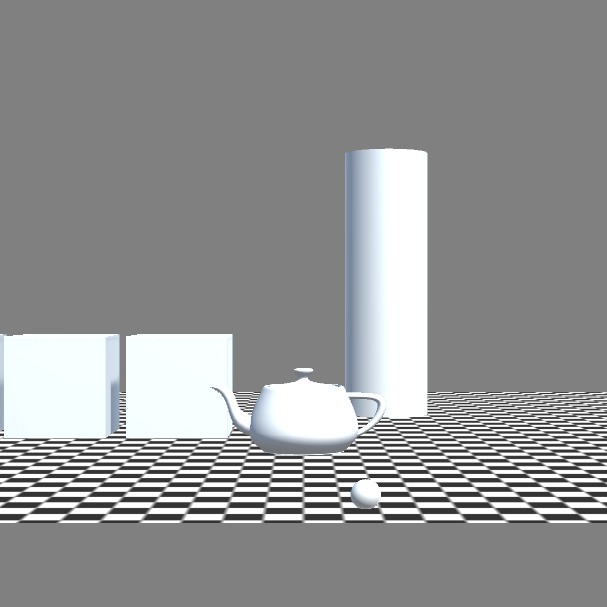}
	\end{tabular}
\end{center}

\end{document}